\documentclass{article}
\usepackage{amssymb}
\usepackage{amsmath}

\begin{document}

\title{One-mode quantum Gaussian channels}
\author{A. S. Holevo}
\maketitle

\begin{abstract}A classification of one-mode Gaussian channels is
given up to canonical unitary equivalence. A complementary to the
quantum channel with additive classical Gaussian noise is
described providing an example of one-mode Gaussian channel which
is neither degradable nor anti-degradable.
\end{abstract}

\section{The canonical form}

For a mathematical framework of linear Bosonic system used in this
note the reader is referred to \cite{hw}. Consider Bosonic system
with one degree of freedom described by the canonical observables
$Q,P$ satisfying the Heisenberg canonical commutation relations
(CCR)
\begin{equation}
\lbrack Q,P]=iI.  \label{Hei}
\end{equation}

Let $Z$ be the two-dimensional symplectic space, i. e. the linear space of
vectors $z=[x,y]$ with the symplectic form
\begin{equation}
\Delta (z,z^{\prime })=x^{\prime }y-xy^{\prime }.  \label{sympl-form}
\end{equation}
A basis $e,h$ in $Z$ is symplectic if $\Delta (e,h)=1,$ i.e. if the area of
the oriented parallelogram based on $e,h$ is equal to 1. A linear
transformation $T$ in $Z$ is symplectic if it maps a symplectic basis into
symplectic basis, or equivalently
\begin{equation*}
\Delta (Tz,Tz^{\prime })=\Delta (z,z^{\prime });\quad z,z^{\prime }\in Z.
\end{equation*}

Let $V(z)=\exp \,i(xQ+yP)$ be the unitary Weyl operators in a
Hilbert space $ \mathcal{H}$ satisfying the CCR
\begin{equation*}
V(z)V(z^{\prime })=\exp [\frac{i}{2}\Delta (z,z^{\prime })]V(z+z^{\prime })
\label{weyl}
\end{equation*}
formally equivalent to (\ref{Hei}).  For any symplectic transformation $T$
in $Z$ there is a canonical unitary transformation $U_{T}$ in $\mathcal{H}$
such that
\begin{equation*}
U_{T}^{\ast }V(z)U_{T}=V(Tz).
\end{equation*}

An arbitrary Gaussian channel $\Phi $ in $\mathfrak{B}\left(
\mathcal{H} \right) $ has the following action on the Weyl operators
(we use the dual channel $\Phi ^{\ast }$ in Heisenberg picture)
\begin{equation}
\Phi ^{\ast }(V(z))=V(Kz)f(z),  \label{linbos}
\end{equation}
where $K$ is a linear transformation in $Z$ while $f(z)$ is a Gaussian
characteristic function satisfying the condition that for arbitrary finite
collection $\left\{ z_{r}\right\} \subset Z$ the matrix with the elements
\begin{equation}
f(z_{r}-z_{s})\,\exp \bigl(-\frac{i}{2}\Delta
(z_{r},z_{s})+\frac{i}{2} \Delta (Kz_{r},Kz_{s})\bigr)  \label{pd}
\end{equation}
is positive definite. Considering the function $f(z),$ we can always
eliminate the linear terms by a canonical transformation and assume
that $ f(z)=\exp \left[ -\frac{1}{2}\alpha (z,z)\right] ,$ where
$\alpha $ is a quadratic form. Then (\ref{pd}) is equivalent to
positive definiteness of the matrix with the elements
\begin{equation}
\alpha (z_{r},z_{s})-\frac{i}{2}\Delta (z_{r},z_{s})+\frac{i}{2}\Delta
(Kz_{r},Kz_{s}).  \label{pdqf}
\end{equation}

Motivated by \cite{cg}, \cite{sew}, we are interested in the simplest form
of one-mode Gaussian channel which can be obtained by applying suitable
canonical unitary transformations to the input and the output of the
channel:
\begin{equation*}
\Phi ^{\ast \prime }\left[ V(z)\right] =U_{T_{1}}^{\ast }\Phi ^{\ast }\left[
U_{T_{2}}^{\ast }V(z)U_{T_{2}}\right] U_{T_{1}}
\end{equation*}
i.e.
\begin{equation*}
\Phi ^{\ast \prime }\left[ V(z)\right] =V(T_{1}KT_{2}z)f(T_{2}z).
\end{equation*}

\textbf{Theorem.} Let $e,h$ be a symplectic basis; depending on the value
\begin{equation*}
A)\quad \Delta (Ke,Kh)=0;\quad B)\quad \Delta (Ke,Kh)=1;\quad
\end{equation*}
\begin{equation*}
C)\quad \Delta (Ke,Kh)=k^{2}>0,k\neq 1;\quad D)\quad \Delta (Ke,Kh)=-k^{2}<0
\end{equation*}
there are symplectic transformations $T_{1},T_{2},$ such that $\Phi ^{\ast
\prime }$ has the form (\ref{linbos}) with
\begin{eqnarray*}
A_1)\quad K\left[ x,y\right]  &=&\left[0,0\right] ; \\
f(z) &=&\exp \left[ -\frac{1}{2}\bigl(\
N_{0}+\frac{1}{2}\bigr)\,\left(
x^{2}+y^{2}\right) \right] ;\quad N_{0}\geq 0; \\
A_2)\quad K\left[ x,y\right]  &=&\left[x,0\right] ;  \\
f(z) &=&\exp \left[ -\frac{1}{2}\bigl(\ N_{0}+\frac{1}{2}\bigr)\,\left(
x^{2}+y^{2}\right) \right] ; \\
B_{1})\quad K\left[ x,y\right]  &=&\left[ x,y\right] ; \\
f(z) &=&\exp \left[ -\frac{1}{4}x^{2}\right] ; \\
B_{2})\quad K\left[ x,y\right]  &=&\left[ x,y\right] ; \\
f(z) &=&\exp \left[ -\frac{1}{2}\ N_{c}\,\left( x^{2}+y^{2}\right) \right]
;\quad N_{c}\geq 0; \\
C)\quad K\left[ x,y\right]  &=&\left[ kx,ky\right] ;\quad k>0,k\neq 1; \\
\quad f(z) &=&\exp \left[ -\frac{|k^{2}-1|}{2}\bigl(\
N_{0}+\frac{1}{2}\bigr)
\,\left( x^{2}+y^{2}\right) \right] ; \\
D)\quad K\left[ x,y\right]  &=&\left[ kx,-ky\right] ;\quad k>0;\quad  \\
f(z) &=&\exp \left[ -\frac{(k^{2}+1)}{2}\bigl(\
N_{0}+\frac{1}{2}\bigr) \,\left( x^{2}+y^{2}\right) \right] .
\end{eqnarray*}

\textit{Proof. }

A) $\Delta (Ke,Kh)=0.$ In this case $\Delta (Kz,Kz^{\prime })\equiv 0$ and
either $K=0$ or $K$ has rank one. Then positive definiteness of (\ref{pdqf})
is just the condition on the correlation function of a quantum Gaussian
state. As follows from Williamson theorem, see e. g. \cite{hw}, there is a
symplectic transformation $T_{2}$ such that
\begin{equation*}
\alpha (T_{2}z,T_{2}z)=\bigl(N_{0}+\frac{1}{2}\bigr)\,\left(
x^{2}+y^{2}\right) .
\end{equation*}
In the case $K=0$ we just have $A_1)$.

 Let $K$ have rank one, then $
KT_{2}$ has rank one and there is a vector $e^{\prime }$ such that
$KT_{2} \left[ x,y\right] = e^{\prime }.$ Then there is a
symplectic transformation $T_{1}$ such that $T_{1}e^{\prime
}=[1,0]$ and hence $T_{1}KT_{2}\left[ x,y\right] =\left[
k_{1}x+k_{2}y,0\right] .$ By making a rotation $T_{2}'$ which
leaves $\alpha$ unchanged, we can transform this vector to $\left[
k_{1}'x,0\right]$ with $k_{1}'\neq 0$, and then by a symplectic
scaling (squeezing) $T_{1}'$ we can transform to the case $A_2)$.

B,C) $\Delta (Ke,Kh)=k^{2}>0.$ Then $T=k^{-1}K$ is symplectic
transformation and $\Delta (Kz,Kz^{\prime })=k^{2}\Delta
(z,z^{\prime }).$ Let $ T_1=(TT_2)^{-1}$ where $T_2$ will be chosen
later, so that $T_1KT_2=kI$. If $ k=1$ then positive definiteness of
(\ref{pdqf}) is just positive definiteness of $\alpha .$ The
transformation $T_2$ is chosen as follows:

In the case B) $k=1$ and positive definiteness of (\ref{pdqf}) is just
positive definiteness of $\alpha .$ Then we have the subcases:

B$_{2})$ If $\alpha $ is nondegenerate then by Williamson theorem there is a
symplectic transformation $T_{2}$ such that
\begin{equation*}
\alpha (T_{2}z,T_{2}z)=\ N_{c}\,\left( x^{2}+y^{2}\right) ,
\end{equation*}
where $N_{c}>0.\,\ $Also if $\alpha =0$ we have a similar formula
with $ N_{c}=0.$

B$_{1})$ On the other hand, if $\alpha $ is degenerate of rank
one, that is $\alpha (z,z)=\left( k_{1}x+k_{2}y\right)^2$ for some
$k_{1},k_{2}$ simultaneously not equal to zero, then for arbitrary
$ N_{c}>0$ there is a symplectic transformation $T_{2}$ such that
\begin{equation*}
\alpha (T_{2}z,T_{2}z)=\ N_{c}\,y^{2}
\end{equation*}
in particular, we can take $ N_{c}=\frac{1}{2}$.

 In the case C) $k\not=1,$ and positive
definiteness of (\ref{pdqf}) implies that $\alpha /|k^{2}-1|$ is
correlation function of a quantum Gaussian state, hence .
\begin{equation*}
\alpha (T_{2}z,T_{2}z)=|k^{2}-1|\bigl(\ N_{0}+\frac{1}{2}\bigr)\,\left(
x^{2}+y^{2}\right)
\end{equation*}
with $N_{0}\geq 0.$

D) $\Delta (Ke,Kh)=-k^{2}<0.$ Then $T=k^{-1}K$ is antisymplectic
transformation $\left( \Delta (Tz,Tz^{\prime })=-\Delta
(z,z^{\prime })\right) $ and $\Delta (Kz,Kz^{\prime
})=-k^{2}\Delta (z,z^{\prime }).$ Similarly to the case B,C) we
obtain
\begin{equation*}
\alpha (T_{2}z,T_{2}z)=|k^{2}+1|\bigl(\ N_{0}+\frac{1}{2}\bigr)\,\left(
x^{2}+y^{2}\right)
\end{equation*}
with $N_{0}\geq 0.$\ Letting $\ T_{1}=\Lambda \left( TT_{2}\right)
^{-1},$ where $\Lambda \left[ x,y\right] =\left[ x,-y\right] ,$ we
obtain the first equation in D). Note that $T_{1}$ is symplectic
since both $T$ and $\Lambda $ are antisymplectic. $\square $

\section{Description in terms of open Bosonic system}

As explained in \cite{hw}, a Gaussian channel $\Phi $ can be
dilated to a linear dynamics (i. e. symplectic transformation) of
open Bosonic system described by $Q,P$ and ancillary canonical
variables $q,p,\dots $ in a Gaussian state $\rho _{0};$ moreover
this linear dynamics provides also a description of the channel
$\Phi _{E}$ mapping initial state of the system $ Q,P$ into the
final state of the ancilla (environment) $q,p,\dots $; in the case
where the state $\rho _{0}$ is pure, $\Phi _{E}$ is just the
complementary channel $\tilde{\Phi}$ in the sense of \cite{ds},
\cite{h}, \cite{mr}, which is determined by $\Phi $ up to unitary
equivalence. In the case of arbitrary $\rho _{0},$ following
\cite{g}, we will call the channel $\Phi _{E}$ weak complementary.

Let us give this description in each of the cases of the previous
Section.

A$_1)$ This is completely depolarizing channel
\begin{eqnarray*}
Q &\rightarrow &q \\
P &\rightarrow &p
\end{eqnarray*}
where $q,p$ are in the quantum thermal state  $\rho _{0}$ with
mean number of quanta $N_{0}.$ Its weak complementary is the ideal
channel $\mathrm{Id}$.

A$_2)$ The linear transformation of the mode canonical variables
is
\begin{eqnarray}\label{a2}
Q &\rightarrow &Q+q \\
P &\rightarrow &p,\nonumber
\end{eqnarray}
where $q,p$ are in the quantum thermal state  $\rho _{0}$ with
mean number of quanta $N_{0}.$ The signal in this channel is
``classical'' in that it has commuting components, so its quantum
capacity must be equal to zero. In fact, it can been shown
\cite{g} that this channel is anti-degradable\footnote{Channel
$\Phi$ is \textit{degradable} \cite{ds} if $\tilde{\Phi}=T\circ
\Phi$ for some channel $T$. It is called \textit{anti-degradable}
\cite{cg} if $\Phi=T'\circ \tilde{\Phi}$ for some channel $T'$.},
hence the conclusion follows. Weak complementary of this channel
is described by the transformation
\begin{eqnarray}\label{wca2}
q &\rightarrow &Q \\
p &\rightarrow &P-p ,\nonumber
\end{eqnarray}
where $p $ can be regarded as a classical real Gaussian variable
with variance $N_0+\frac{1}{2}$.

B$_{1})$ The equation of the channel has the form (\ref{wca2})
where $p$ has variance $\frac{1}{2}$, so that the mode $q,p$ is in
pure (vacuum) state, and the complementary channel is given by
(\ref{a2}), so that by the previous remark, the channel is
degradable. Its quantum capacity is infinite as can be seen from
the lower bound given by the Gaussian coherent information
resulting from the Gaussian input states with
$\sigma_P^2=\mathsf{D}P\to\infty$. In more detail, denoting the
channel given by the equation (\ref{wca2}) by $\Phi$ and its
complementary -- by $\tilde{\Phi}$ we have for the coherent
information $J(\rho ,\Phi )=H(\Phi [\rho ])-H( \tilde{\Phi}[\rho
])$. Now let $\rho$ be the Gaussian state with
$\sigma_Q^2=\mathsf{D}Q, \sigma_P^2=\mathsf{D}P$ and zero means
and covariance, then $\Phi [\rho ]$ has the correlation matrix
\begin{equation*}\left[
  \begin{pmatrix}
    \sigma_Q^2 & 0 \\
    0 & \sigma_P^2+\frac{1}{2} \
  \end{pmatrix}\right],
\end{equation*}
while $\tilde{\Phi}[\rho ]$ has the correlation matrix
\begin{equation*}\left[
  \begin{pmatrix}
    \sigma_Q^2+\frac{1}{2} & 0 \\
    0 & \frac{1}{2} \
  \end{pmatrix}\right].
\end{equation*}
By a formula from \cite{hw},
\begin{eqnarray*}
H(\Phi [\rho ]) &= & g\left(\sqrt{\sigma_Q^2\left(\sigma_P^2+\frac{1}{2}\right)}-\frac{1}{2}\right),\\
H( \tilde{\Phi}[\rho ]) &= &
g\left(\sqrt{\left(\sigma_Q^2+\frac{1}{2}\right)\frac{1}{2}}-\frac{1}{2}\right),
\end{eqnarray*}
where $g(x)=(x+1)\log (x+1)-x\log x$. In particular, taking states
with $\sigma_Q^2\sigma_P^2\to\infty,\sigma_Q^2\to 0$, we have
$J(\rho ,\Phi )\to\infty$.

 B$_{2}$) Channel with additive complex Gaussian
noise of intensity $N_{c}$
\begin{eqnarray*}
Q &\rightarrow &Q+\xi \\
P &\rightarrow &P+\eta
\end{eqnarray*}
to be described in detail in Sec. 3.

C) Attenuation/amplification channel with coefficient $k$ and
quantum noise with mean number of quanta $N_{0}.$ In the attenuation
case ($k<1$) \ the equation of the channel is
\begin{eqnarray*}
Q &\rightarrow &kQ+\sqrt{1-k^{2}}q \\
P &\rightarrow &kP+\sqrt{1-k^{2}}p,
\end{eqnarray*}
where  $q,p$ are in the quantum thermal state $\rho _{0}$ with mean
number of quanta $N_{0}.$ The weak complementary is given by the
equations
\begin{eqnarray*}
q &\rightarrow &\sqrt{1-k^{2}}Q-kq \\
p &\rightarrow &\sqrt{1-k^{2}}P-kp,
\end{eqnarray*}
and is again an attenuation channel (with coefficient
$k'=\sqrt{1-k^{2}}$) see \cite{h}, \cite{cg}.

\bigskip In the amplification case ($k>1$) we have
\begin{eqnarray*}
Q &\rightarrow &kQ+\sqrt{k^{2}-1}q \\
P &\rightarrow &kP-\sqrt{k^{2}-1}p,
\end{eqnarray*}
with the weak complementary
\begin{eqnarray*}
q &\rightarrow &\sqrt{k^{2}-1}Q+kq \\
p &\rightarrow &-\sqrt{k^{2}-1}P+kp,
\end{eqnarray*}
see the case D).

Alternatively, by introducing $N_{c}=|k^{2}-1|N_{0}\ge 0$, the
case C) is the same as attenuation/amplification channel with
vacuum ancilla state and additive classical Gaussian noise of
intensity $N_{c}$ considered in \cite{hw}, where one-shot Gaussian
coherent information for this channel was computed. From \cite{cg}
it follows that in the case $N_{0}=0$ the channel C) is
degradable, hence the coherent information is subadditive in this
case by \cite{ds} and the quantum capacity is equal to the maximum
of one-shot coherent information. Moreover, it was observed in
\cite{wpgg} that the one-shot coherent information is concave for
degradable channels, hence it is maximized by a Gaussian state. To
sum up, the quantum capacity of the attenuation/amplification
channel with $N_{0}=0$ and coefficient $k$ is equal to the
maximized one-shot Gaussian coherent information
$\max\{0,\log\frac{k^2}{|k^2-1|}\}$, an expression following from
\cite{hw}. The case C) with $N_{0}>0$ as well as B$_2$) remain
open questions.

D) The channel equation is
\begin{eqnarray*}
Q &\rightarrow &kQ+\sqrt{k^{2}+1}q \\
P &\rightarrow &-kP+\sqrt{k^{2}+1}p,
\end{eqnarray*}
which is the same as the weak complementary to amplification
channel with coefficient $k^{\prime }=\sqrt{k^{2}+1}$ and quantum
noise with mean number of quanta $N_{0},$ see \cite{h} ,
\cite{cg}. It was shown in \cite{cg} that this channel is
anti-degradable, hence has zero quantum capacity.

\section{Quantum signal plus classical Gaussian noise}

\bigskip Here we give explicit construction of complementary channel in case
B$_{2})$ based on corrected and improved presentation of Appendix
B in \cite{hw}.

Introducing the one mode annihilation operator $a=\frac{1}{\sqrt{2}}(Q+iP)$
consider the transformation
\begin{equation}
a^{\prime }=a+\zeta ,  \label{splusnoise}
\end{equation}
where $\zeta= \frac{1}{\sqrt{2}}(\xi+i\eta)$ is a complex Gaussian
random variable with zero mean and variance $N_{c}$. This means that
in the plane of the complex variable $z= \frac{1}{\sqrt{2}}(x+iy)$
it has the probability distribution $\ \mu _{N_{c}}(d^{2}z)$ with
the density
\begin{equation*}
p(z)=\left( \pi N_{c}\right) ^{-1}\exp \left( -|z|^{2}/N_{c}\right)
\end{equation*}

This transformation generates the channel
\begin{equation*}
\Phi^{\ast }[f(a,a^{\dagger })]=\int f(a+z,(a+z)^{\dagger
})p(z)d^{2}z,
\end{equation*}
in the Heisenberg picture, while the channel in the Schr\"odinger
picture can be described by the formula
\begin{equation}
\Phi[\rho ]=\int D(z)\rho D(z)^{\ast }p(z)d^{2}z,  \label{spn}
\end{equation}
where $D(z)=\exp \left( za^{\dagger }-\bar{z}a\right) =\exp i\left(
yQ-xP\right)=V(-Jz) $ is the displacement operator. Here
$J[x,y]=[-y,x]$ is the operator of complex structure. The channel
describes the quantum mode $Q,P$ in the additive classical Gaussian
environment $\zeta$ i. e. the case B$_2$) of Sec. 1.

Here we construct the ancilla representation of the channel as unitary
evolution of the mode $Q,P$ interacting with quantum environment (ancilla),
and find the complementary channel for $\Phi $. We also argue that for
certain values of $N_{c}$ both channels have positive quantum capacities and
hence are neither degradable nor anti-degradable, cf. \cite{cg}.

First we need to extend the classical environment to a quantum
system in a pure state. Consider the ancilla Hilbert space
$\mathcal{H}_{E}=L^{2}(\mu _{N_{c}})$ with the vector $|\Psi
_{0}\rangle $ given by the function identically equal to 1. The
tensor product $\mathcal{H}\otimes \mathcal{H} _{E}$ can be realized
as the space $L_{\mathcal{H}}^{2}(\mu _{N_{c}})$ of $ \mu
_{N_{c}}$-square integrable functions $\psi (z)$ with values in
$\mathcal{ H}$. Define the unitary operator $U$ in
$\mathcal{H}\otimes \mathcal{H}_{E}$ by
\begin{equation}
(U\psi )(z)=D(z)\psi (z).  \label{dyn}
\end{equation}
Then
\begin{equation*}
\Phi \lbrack \rho ]=\mathrm{Tr}_{\mathcal{H}_{E}}U\left( \rho \otimes |\Psi
_{0}\rangle \langle \Psi _{0}|\right) U^{\ast },
\end{equation*}
while the complementary channel is
\begin{equation*}
\Phi _{E}[\rho ]=\tilde{\Phi}[\rho ]=\mathrm{Tr}_{\mathcal{H}}U\left( \rho
\otimes |\Psi _{0}\rangle \langle \Psi _{0}|\right) U^{\ast }.
\end{equation*}
This means that $\tilde{\Phi}[\rho ]$ is an integral operator in $L^{2}(\mu
_{N_{c}})$ with the kernel
\begin{equation*}
K(z,z^{\prime })=\mathrm{Tr}D(z)\rho D(z^{\prime })^{\ast }.
\end{equation*}
In case of Gaussian state $\rho _{N}$ with the characteristic function
\begin{equation}
\mathrm{Tr}\rho _{N}D(z)=\exp (-(N+1/2)|z|^{2})  \label{ro0}
\end{equation}
we have
\begin{equation*}
K(z,z^{\prime })=\exp (i\Im \bar{z}^{\prime }z-(N+1/2)|z-z^{\prime }|^{2}).
\end{equation*}

Let us define the canonical observables in $L^{2}(\mu _{N_{c}}).$ For this
consider the unitary Weyl operators $V(z_{1},z_{2})$ in $L^{2}(\mu _{N_{c}})$
by
\begin{eqnarray}
(V(z_{1},z_{2})\psi )(z) &=&\psi (z+z_{2})  \label{we} \\
&&\times \exp \left[ i2\Re
\overline{z_{1}}(z+\frac{z_{2}}{2})-\frac{1}{N_{c} }\Re
\overline{z_{2}}(z+\frac{z_{2}}{2})\right] . \nonumber
\end{eqnarray}
The operators $V(z_{1},z_{2})$ satisfy the Weyl-Segal CCR
\begin{equation}
V(z_{1},z_{2})V(z_{1}^{\prime },z_{2}^{\prime })=\exp [\frac{i}{2}\Delta
((z_{1},z_{2}),(z_{1}^{\prime },z_{2}^{\prime }))]V(z_{1}+z_{1}^{\prime
},z_{2}+z_{2}^{\prime }),  \label{ccr}
\end{equation}
with two degrees of freedom with respect to the symplectic form
\begin{equation*}
\Delta ((z_{1},z_{2}),(z_{1}^{\prime },z_{2}^{\prime }))=2\Re \left(
\bar{z} _{1}^{\prime }z_{2}-\bar{z}_{1}z_{2}^{\prime }\right)
=x_{1}^{\prime }x_{2}-x_{1}x_{2}^{\prime }+y_{1}^{\prime
}y_{2}-y_{1}y_{2}^{\prime },
\end{equation*}
where $z_{j}=\frac{1}{\sqrt{2}}(x_{j}+iy_{j});j=1,2.$ By the general theory
of CCR (see e.g. \cite{hw}) one has
\begin{equation*}
V(z_{1},z_{2})=\exp \,i(x_{1}q_{1}+y_{1}q_{2}+x_{2}p_{1}+y_{2}p_{2}),
\end{equation*}
where the canonical observables $q_{j},p_{j};j=1,2,$ satisfy the Heisenberg
CCR
\begin{equation*}
\lbrack q_{j},p_{k}]=i\delta _{jk}I,\;\;[q_{j},q_{k}]=0,\;\;[p_{j},p_{k}]=0,
\end{equation*}
and commute with $Q,P.$ By differentiating (\ref{we}) with respect
to $ x_{1},x_{2}$ at zero point, we obtain that
\begin{equation}
(q_{1}\psi )(z)=x\psi (z);\quad (q_{2}\psi )(z)=y\psi (z).  \label{qq}
\end{equation}
The state of ancilla is pure and given by the function in $L^{2}(\mu
_{N_{c}})$ identically equal to one. It has the characteristic function
\begin{eqnarray*}
&&\int (V(z_{1},z_{2})1)(z)p(z)d^{2}z \\
&=&\left( \pi N_{c}\right) ^{-1}\int \exp \left[ i2\Re
\overline{z_{1}}(z+ \frac{z_{2}}{2})-\frac{1}{N_{c}}\Re
\overline{z_{2}}(z+\frac{z_{2}}{2})-
\frac{|z|^{2}}{N_{c}}\right] d^{2}z \\
&=&\exp \left[ -N_{c}|z_{1}|^{2}-\frac{1}{4N_{c}}|z_{2}|^{2}\right] ,
\end{eqnarray*}
which means that it is pure Gaussian with $q_{j},p_{j};j=1,2$ having zero
means, zero covariances and the variances
\begin{equation*}
\mathsf{D}q_{1}=\mathsf{D}q_{2}=N_{c};\quad
\mathsf{D}p_{1}=\mathsf{D}p_{2}= \frac{1}{4N_{c}}.
\end{equation*}

The Heisenberg dynamics of the Weyl operators of system+ancilla is given by
the equations
\begin{eqnarray}
D(z_{0}) &\rightarrow &D(z)^{\ast }D(z_{0})D(z)=\exp 2i\Im (\bar{z}
z_{0})D(z_{0}),  \label{d1} \\
V(z_{1},z_{2}) &\rightarrow &D(z)^{\ast }V(z_{1},z_{2})D(z)=\exp
i\Im (\bar{z }z_{2})D(z_{2})V(z_{1},z_{2}),  \label{d2}
\end{eqnarray}
where the first is the CCR for the displacement operators, while
the second follows from (\ref{dyn}), (\ref{we}). By
differentiating these relations with respect to
$x_{j},y_{j};j=0,1,2,$ and taking into account (\ref{qq}) we
obtain the Heisenberg equations for the canonical observables
\begin{eqnarray*}
Q &\rightarrow &Q+q_{1}, \\
P &\rightarrow &P+q_{2}, \\
q_{1} &\rightarrow &q_{1}, \\
p_{1} &\rightarrow &p_{1}-P-q_{2}/2, \\
q_{2} &\rightarrow &q_{2}, \\
p_{2} &\rightarrow &p_{2}+Q+q_{1}/2.
\end{eqnarray*}

Together with the ancilla state described above, the first two
equations are the same as the equation (\ref{splusnoise})
determining the channel $\Phi $, while the last four equations give
the action of the complementary channel $ \tilde{\Phi}$ in terms of
the canonical observables.

The characteristic function of the output ancilla state $\tilde{\Phi}[\rho ]$
is
\begin{equation}
\mathrm{Tr}\tilde{\Phi}[\rho ]V(z_{1},z_{2})=\mathrm{Tr}\rho
D(z_{2})\exp \left[ -\left( N_{c}|z_{1}|^{2}+N_{c}\Im
\bar{z}_{1}z_{2}+\frac{N_{c}{}^{2}+1 }{4N_{c}}|z_{2}|^{2}\right)
\right] .  \label{chfch}
\end{equation}
\textit{Proof of (\ref{chfch}).} By using (\ref{d2}) we have
\begin{eqnarray*}
\mathrm{Tr}\tilde{\Phi}[\rho ]V(z_{1},z_{2}) &=&\int \mathrm{Tr}D(z)^{\ast
}V(z_{1},z_{2})D(z)\rho p(z)d^{2}z \\
&=&\int \exp i\Im (\bar{z}z_{2})\mathrm{Tr}\rho
D(z_{2})(V(z_{1},z_{2})1)(z)p(z)d^{2}z
\end{eqnarray*}
which is equal to $\mathrm{Tr}\rho D(z_{2})$ multiplied by the integral
\begin{equation*}
\int \exp \left[ i\Im (\bar{z}z_{2})+i2\Re
\overline{z_{1}}(z+\frac{z_{2}}{2} )-\frac{1}{N_{c}}\Re
\overline{z_{2}}(z+\frac{z_{2}}{2})-\frac{|z|^{2}}{N_{c} }\right]
\frac{d^{2}z}{\pi N_{c}}.
\end{equation*}
By introducing the change of variables $z+\frac{z_{2}}{2}=w,$ we obtain that
this is equal to
\begin{eqnarray*}
&&\exp \left[ -\frac{|z_{2}|^{2}}{4N_{c}}\right] \int \exp \left[
i\Im (\bar{ w}z_{2})+i2\Re
\overline{z_{1}}w-\frac{|w|^{2}}{N_{c}}\right] \frac{d^{2}w}{
\pi N_{c}} \\
&=&\exp \left[ -\frac{|z_{2}|^{2}}{4N_{c}}\right] \int \exp \left[
i2\Re \bar{w}\left( z_{1}-i\frac{z_{2}}{2}\right)
-\frac{|w|^{2}}{N_{c}}\right] \frac{d^{2}w}{\pi N_{c}}
\end{eqnarray*}
where the last integral is just the characteristic function of the complex
Gaussian distribution, equal to
\begin{equation*}
\exp \left[ -N_{c}\left\vert z_{1}-i\frac{z_{2}}{2}\right\vert
^{2}\right] =\exp \left[ -N_{c}\left( |z_{1}|^{2}+\Im
\bar{z}_{1}z_{2}+\frac{1}{4} |z_{2}|^{2}\right) \right]
\end{equation*}
whence (\ref{chfch}) follows. $\square $

In the case of Gaussian $\rho =\rho _{N}$ given by (\ref{ro0}), we obtain
\begin{equation*}
\mathrm{Tr}\tilde{\Phi}[\rho _{N}]V(z_{1},z_{2})=\exp \left[ -\left(
N_{c}|z_{1}|^{2}+N_{c}\Im \bar{z}_{1}z_{2}+\frac{D^{2}}{4N_{c}}
|z_{2}|^{2}\right) \right] ,
\end{equation*}
where $D=\sqrt{(N_{c}+1)^{2}+4N_{c}N},$ so that
$\frac{D^{2}}{4N_{c}}=\frac{1
}{4}(N_{c}+\frac{1}{N_{c}})+(N+\frac{1}{2})$, which is Gaussian
characteristic function with the correlation matrix
\begin{equation*}
\alpha _{E}^{\prime }=\left[
\begin{array}{cccc}
N_{c} & 0 & 0 & N_{c}/2 \\
0 & N_{c} & -N_{c}/2 & 0 \\
0 & -N_{c}/2 & \frac{D^{2}}{4N_{c}} & 0 \\
N_{c}/2 & 0 & 0 & \frac{D^{2}}{4N_{c}}
\end{array}
\right]
\end{equation*}
Thus, with the commutation matrix of the ancilla
\begin{equation*}
\Delta _{E}=\left[
\begin{array}{cccc}
0 & 0 & 1 & 0 \\
0 & 0 & 0 & 1 \\
-1 & 0 & 0 & 0 \\
0 & -1 & 0 & 0
\end{array}
\right]
\end{equation*}
we obtain
\begin{equation*}
\ \Delta _{E}^{-1}\alpha _{E}^{\prime }=\left[
\begin{array}{cccc}
0 & N_{c}/2 & -\frac{D^{2}}{4N_{c}} & 0 \\
-N_{c}/2 & 0 & 0 & -\frac{D^{2}}{4N_{c}} \\
N_{c} & 0 & 0 & N_{c}/2 \\
0 & N_{c} & -N_{c}/2 & 0
\end{array}
\right] .
\end{equation*}
which has the eigenvalues $\lambda _{\epsilon}=\pm
\frac{i}{2}\left( N_{c}\pm D\right); \epsilon=1,2,3,4.$

Let the input state be Gaussian $\rho _{N}$ given by (\ref{ro0}),
then the entropy of $\rho _{N}$ is $H(\rho _{N})=g(N)$. The output
state has the entropy

\begin{equation}
H(\Phi [\rho _{N}])=g(N+N_{c}),  \label{out-1mode}
\end{equation}
while the exchange entropy is
\begin{eqnarray}
H(\rho _{N},\Phi ) &=&H(\tilde{\Phi}[\rho _{N}])  \label{exch-1mode} \\
&=&\frac{1}{2}\sum_{\epsilon=1}^4g\left(
|\lambda_{\epsilon}|-\frac{1}{2}\right) \nonumber \\ &=&g\left(
\frac{D+N_{c}-1}{2}\right) +g\left(
\frac{D-N_{c}-1}{2}\right),\nonumber
\end{eqnarray}
see \cite{hw} for detail.

Introducing the coherent information $J(\rho ,\Phi )=H(\Phi [\rho
])-H( \tilde{\Phi}[\rho ]),$ we have lower bounds for the quantum
capacities $ Q(\Phi ),$ $Q(\tilde{\Phi}):$
\begin{eqnarray*}
\sup_{\rho }J(\rho ,\Phi ) &\leq &Q(\Phi ), \\
\sup_{\rho }\left[ -J(\rho ,\Phi )\right]  &\leq &Q(\tilde{\Phi}).
\end{eqnarray*}
Therefore if $J(\rho ,\Phi )$ accepts the values of different signs,
both capacities are positive, hence by using results from \cite{cg},
the channel $ \Phi $ is neither anti-degradable nor degradable.
Consider the function
\begin{equation*}
F(N)=J(\rho _{N},\Phi )=g(N+N_{c})-g\left( \frac{D+N_{c}-1}{2}\right)
-g\left( \frac{D-N_{c}-1}{2}\right) ,\quad N\geq 0.
\end{equation*}
We have $F(0)=0;$ by using the asymptotic $g(x)\simeq \log ex,$ $
x\rightarrow \infty ,$ we obtain
\begin{equation*}
\lim_{N\rightarrow \infty }F(N)=-\log eN_{c},
\end{equation*}
hence $\lim_{N\rightarrow \infty }F(N)=0$ if $N_{c}=e^{-1}.$ Considering the
graph of $F(N)$ for $N_{c}=0.99e^{-1},$ one can see that $F(N)$ has a small
negative peak for small positive $N,$ while $\lim_{N\rightarrow \infty
}F(N)=-\log 0.99>0.$ Thus $F(N)$ and hence $J(\rho ,\Phi )$ accepts the
values of different signs, and the channel $\Phi $ is neither
anti-degradable nor degradable.

\textbf{Acknowledgements.} This research was partially supported
by RFBR grant 06-01-00164-a and the scientific program
"Theoretical problems of modern mathematics". The paper was
completed when the author was the Leverhulme Visiting Professor at
CQC, DAMTP, University of Cambridge. The author is grateful to V.
Giovannetti, Y. M. Suhov and M. Shirokov for discussions.

\end{document}